\documentstyle[11pt,amsfonts,amssymb,fullpage,here]{article}
\input{psfig}
\begin{document}

\begin{titlepage}
\begin{center}
{\large
XIX INTERNATIONAL SYMPOSIUM ON LEPTON AND PHOTON INTERACTION AT HIGH
ENERGY \\

Stanford University, August 9-14, 1999} \\[4cm]

{\Large L.G.Landsberg\\
State Research Center, Institute for High Energy Physics,\\ 
Protvino, Moscow region, Russia, 142284 \\[2cm]

{\bf
SEARCH FOR EXOTIC BARYONS WITH HIDDEN \\
STRANGENESS IN THE EXPERIMENTS \\
OF THE SPHINX COLLABORATION$^{*)}$ \\[0.1em]
IN DIFFRACTIVE AND COULOMB PRODUCTION PROCESSES}
}
\end{center}

\vspace*{3.5cm}
\small
\noindent
\rule{5cm}{0.5pt}\\
$^{*)}$The SPHINX Collaboration (IHEP-ITEP):\\
S.V.Golovkin, A.P.Kozhevnikov, V.P.Kubarovsky, V.F.Kurshetsov,
L.G.Landsberg, V.V.Molchanov, V.A.Mukhin, S.V.Petrenko,
D.V.Vavilov, V.A.Victorov\\
{\it{State Research Center, Institute for High Energy Physics, Protvino, 
Russia}}\\
V.Z.Kolganov, G.S.Lomkatsi, A.F.Nilov, V.T.Smolyankin\\
{\it{State Research Center, Institute of Theoretical and Experimental
Physics, Moscow, Russia}}

\end{titlepage}
\begin{abstract}

Experimental results of the SPHINX
Collaboration on studying
 proton diffractive production  processes are presented.
Evidences for new baryon states with masses $\gtrsim~1.8$~GeV were
obtained in  hyperon-kaon effective mass spectra in several reactions.
New data for the diffractive reaction $p+N \rightarrow [\Sigma^oK^+]+N$ at
$E_p=70$~GeV were obtained with partially upgraded SPHINX setup.
The data are in a good agreement with the results of our previous study
of this reaction. In the mass spectrum $M(\Sigma^oK^+)$ a structure
at the threshold region with a mass $\sim 1800$~MeV and a distinct
$X(2000)$ peak with $M=1989\pm 6$~MeV and $\Gamma=91\pm20$~MeV
are observed. Unusual features of the massive $X(2000)$
state (narrow decay width, anomalously large branching ratio for the
decay channel with strange particle emission) make it a serious candidate
for cryptoexotic pentaquark baryon with hidden strangeness $|qqqs \bar s>$.
We also present new results on the narrow threshold structure
$X(1810)$ with $M=1807 \pm 7$~MeV and $\Gamma = 62 \pm 19$~MeV
which is produced in the region of very small $p^2_T<0.01$~GeV$^2$. The
possibility of the Coulomb production mechanism for $X(1810)$ is
discussed.

\end{abstract}

\section{INTRODUCTION}

The last two decades have been marked by a significant increase in
studies in the field of spectroscopy of hadrons, that is, the
particles participating in strong interactions. It has been firmly
established that hadrons are not truly elementary, but composite
particles. Similar to atomic nuclei, which consist of nucleons,
hadrons are bound systems composed of fundamental particles known
as quarks. Quarks are those structural elements that define the variety
of the hadronic matter. In addition to  a fractional
electric charge  and baryon number, different flavors (isospin, strangeness,
charm, etc.) quarks have a  ``strong-interaction charges''
called colors.

The interactions between quarks go through  the exchange of color
virtual particles, gluons, in a similar way as the
interactions between electric charges are effected by exchange of virtual photons.
However, in contrast to neutral photons, which bear no charge and do not
interact directly with each other, gluons are characterized by color ``charges''
and are capable of interacting not only with color quarks but between
themselves, thus forming even bound gluon states~--- glueballs. The
interaction between color quarks and gluons is described by quantum
chromodynamics.

Apparently, color quarks and gluons cannot exist in a free state.
Such a hypothesis is experimentally based on the long-standing unsuccessful
efforts to find free quarks. Therefore, the concept of confinement
(``the quark imprisonment'') arose, according to which only the particles
without color charge can freely exist, the so-called ``colorless'' or
``white'' hadrons. All these hadrons are divided by their  quantum
numbers and their quark content into two large groups: baryons and mesons.
Baryons are characterized by their baryon number B (B=1 for baryons
and B=-1 for antibaryons) and are produced in pairs to conserve the baryon
number of the whole system. Mesons have B=0. Rapid development of hadron
spectroscopy has led to a significant advance in the systematics
of ordinary hadrons with simplest valence quark structure: $q \bar q$
for mesons and $qqq$ for baryons.

Here, we consider only the so-called valence quarks, which determine the hadron
nature and its main characteristics (quantum numbers). According to the
current theoretical perceptions, well confirmed by numerous experiments,
the valence quarks in a hadron are surrounded by a ``cloud'' of
virtual quark-antiquark pairs and gluons, which are constantly emitted
and absorbed by the valence quarks. This ``cloud'' or, as one says now,
the ``sea'' of quark-antiquark pairs and gluons, is physical reality
determining many properties of hadron (for example, space
distribution of its electric charge and magnetic moment, intrinsic
distribution of quark and gluon constituents over momenta, etc.).

A very important question is whether there exist ``colorless''
hadrons with a more complex inner valence structure, such
as multiquark mesons $(M=q q \bar q \bar q)$ and baryons $(B=q q q \bar qq)$,
dibaryons $(D=qqqqqq)$, hybrid states $(M=q \bar q g)$, and glueballs,
which are mesons composed only of gluons $(M=gg,~ggg)$. Of course,
in such new forms of hadronic matter, known as exotic hadrons, effects of the
quark-gluon ``sea'' must also appear in addition to the valence quark and gluon
structure.

The discovery of the exotic hadrons would have a far-reaching consequences for
quantum chromodynamics, for the concept of confinement and for specific models
of hadron structure (lattice, string, and bag models). Detailed discussions of exotic
hadron physics can be found in recent reviews~[1-7].

Exotic hadrons can have anomalous quantum numbers not accessible to
three-quark baryons or quark-antiquark mesons (open exotic states), or
even usual quantum numbers (cryptoexotic states). Cryptoexotic hadrons can be
identified only by their unexpected dynamical properties (anomalously narrow
decay widths, anomalous decay branching ratios and so on).

As is clear from review papers~[1-7], in the last decade searches for
exotic mesons have led to a considerable advance in this field. Several
new states have been observed whose properties cannot be
explained in the framework of naive quark model of ordinary mesons with
$q \bar q$ valence quark structure. These states are serious candidates for
exotic mesons (most of them are of cryptoexotic type).

At the same time the situation for exotic baryons is far
from being clear. There are also some examples of possible
unusual baryon resonances~[8-11], but these data are not sufficiently
precise and are not supported by some new experimental results~[2,12-16].

Extensive studies of the diffractive baryon production and search
for cryptoexotic pentaquark baryons with hidden strangeness
$(B_{\phi}=|qqqs \bar s>$, here $q=u, d$ quarks) are being carried
out by the SPHINX Collaboration at the IHEP accelerator with 70~GeV proton
beam. This program was described in detail in reviews~[2].

The recent data of the SPHINX experiment~[16-25] gave new important evidence
of possible existence of cryptoexotic baryons with hidden strangeness
$X(2050)^+ \rightarrow \Sigma(1385)^oK^+$ and
$X(2000)^+ \rightarrow \Sigma^o K^+$. We shall summarize these data in
Section 3, after a general description of the nature and expected properties
of cryptoexotic baryons, as well as some promising ways for their
production and observation. There were also the SPHINX
results in favor of strong violation of the OZI rule in proton diffractive
dissociation reactions~[26-28] which may be connected with direct strangeness in
the nucleon quark structure.

\section{EXOTIC BARYONS AND POSSIBLE MECHANISMS OF THEIR PRODUCTION}

There arise three main questions tightly connected with the exotic searches
in the SPHINX experiments:

1. How to identify cryptoexotic $B_{\phi}=|qqqs \bar s>$ baryons
without open exotic values of their quantum numbers and how to distinguish
them from several dozens of well-known $N^*$ and $\Delta$ isobars?

2. How to produce the exotic baryons in the most effective way?

3. How to reduce background processes and to make easier the exotic
baryon observation?

We will try to find some qualitative answers to these questions because
of the lack of theoretical models for the description of exotic
hadrons.

\subsection{Properties of Exotic Baryons with Hidden Strangeness}

As has been stated before, cryptoexotic baryons do not have external
exotic quantum numbers, and their complicated internal valence structure can be
established only indirectly by examining their unusual dynamical
properties that are quite different from those of ordinary baryons $|qqq>$.
In this connection, we consider the properties of multiquark baryons with
hidden strangeness $|qqqs \bar s>$.

If such  cryptoexotic baryon structure consists of two color parts
spatially separated by a centrifugal barrier, its decays into the color-singlet
final states
may be suppressed because of a complicated quark rearrangement in decay
processes. The properties of multiquark exotic baryons with the internal color
structure
\begin{equation}
|qqqq \bar q>_{1c} = |(qqq)_{8c} \times (q \bar q)_{8c}>
\end{equation}
(color octet bonds) or
\begin{equation}
|qqqq \bar q>_{1c} = |(qq \bar q)_{\overline{6c}} \times (q  q)_{6c}>
\end{equation}
(color sextet-antisextet bonds) are discussed in~[29-32]. Here, subscripts 1c,
8c, and so on specify representations of the color $SU(3)_c$ group. If the
mass of a nonstrange baryon with hidden strangeness  is above the threshold for
decay modes involving strange particles in final states, the main decay
channels must be of the type
\begin{equation}
B_{\phi}=|qqq s \bar s> \rightarrow YK + k \pi
\end{equation}
($k$=0, 1, ...). Another possibility is associated with the decays
\begin{equation}
B_{\phi} = |qqqs \bar s> \rightarrow \phi N (\eta N; \eta' N) + k \pi.
\end{equation}
which involve the emission of particles with a significant
$s \bar s $ component in their valence-quark structure. It should also be
emphasized that $\eta$ and, particularly, $\eta'$ mesons are strongly
coupled to gluon fields and, hence, to the states with an enriched gluon component.
Therefore, baryon decays of the type $B \rightarrow N \eta,~N \eta'$ may
be specific decay modes for hybrid baryons (see, for example,~[2]).

The nonstrange decays of baryons with hidden strangeness, $B_{\phi} \rightarrow
N + k \pi$, must be suppressed by the OZI rule. Thus, the effective
phase-space factors for the decays of the massive baryons $B_\phi$ would be
significantly reduced because of this OZI suppression (owing to a high
mass threshold for the allowed decays $B_\phi \rightarrow YK$ with respect
to the suppressed decay $B_\phi \rightarrow N \pi,~\Delta \pi$).
The mechanism of quark rearrangement of color clusters in the decays of
particles with complicated inner structure of  type (1) or (2) can further
reduce the decay width of cryptoexotic baryons and make them anomalously narrow
(their widths may become as small as several tens of MeV). Here, theoretical
predictions are rather uncertain. For this reason, only experiments can answer
conclusively the question of whether such narrow baryon  resonances with hidden
strangeness  really exist.

Thus, it is desirable to perform systematic searches for the cryptoexotic
baryons $B_\phi$ with anomalous dynamical features listed below.

(i) The dominant OZI-allowed decay modes of the baryons $B_\phi$ are those
with strange particles in the final states:
\begin{equation}
R(|qqqs \bar s> = BR[|qqqs \bar s> \rightarrow YK]/BR[|qqqs \bar s>
\rightarrow p \pi \pi;~\Delta \pi] \gtrsim 1.
\end{equation}
For ordinary $|qqq>$ isobars $R(\Delta;~N^*)\sim$(several \%)~[33].

ii) The cryptoexotic baryons $B_\phi$ can simultaneously possess a large
mass ($M>1.8-2.0$~GeV) and a narrow decay width
($\Gamma \leq 50-100$~MeV). This is due to a complicated internal color
structure of these baryons, which leads to a significant quark rearrangement of
color clusters in decay processes, and due to a limited
phase space for the OZI-allowed decays $B_\phi \rightarrow YK$. At the same
time, typical decay widths of the well-established
$|qqq>$ isobars with similar masses  are not less
than 300~MeV.

\subsection{Diffractive Production Mechanism and
Search for Exotics}

It was emphasized in a number of studies~[1,2,8,10,30,32,34] that diffractive
production processes featuring Pomeron exchange offer new possibilities
in searches for exotic hadrons. Originally, interest was focused on the
model of Pomeron with a small cryptoexotic $(qq \bar q \bar q)$ component~[30,32].
According to modern concepts, the Pomeron is a multigluon system
owing to which exotic hadrons can be produced in gluon-rich
diffractive processes (see Fig.~1).

\begin{figure}[h]
\centerline{\psfig{file=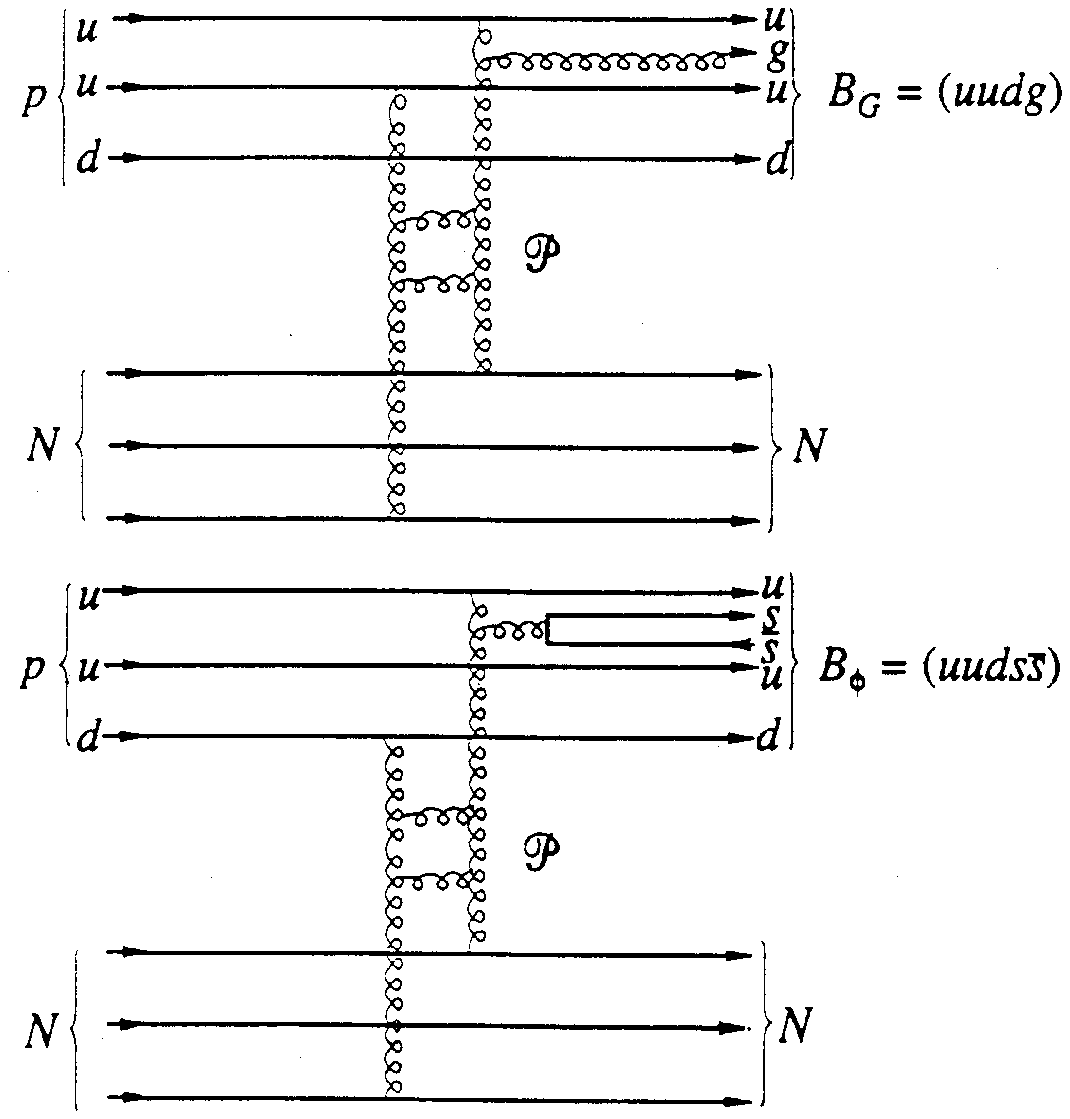,width=10cm}}
\caption{Diagrams for exotic baryon production in the diffraction
processes with the  Pomeron exchange.
The Pomeron {\cal{P}} is a multigluon system.}
\end{figure}

It is apparent that only the states with the same charge and
flavor as those of the primary hadrons can be produced in diffractive
processes. Moreover, there are some additional restrictions on the
spin and parity of the formed hadrons which are stipulated by the
Gribov-Morrison selection rule. According to this rule, the change in parity
$\Delta P$ occurring as a result of the transition from the primary
hadron to the diffractively produced hadronic system, is connected
with the corresponding change in the spin $\Delta J$ through the relation
$\Delta P= (-1)^{\Delta J}$. For example, because of this rule, in the
proton diffractive dissociation  (for proton $J^P=1/2^+$), only baryonic states with
natural sets of quantum numbers $1/2^+,~3/2^-,~5/2^+,~7/2^-$, etc.
can be excited. The Gribov-Morrison selection rule is not a rigorous law and has
an approximate character.

The Pomeron exchange mechanism in diffractive production reactions can induce
the coherent processes on the target nucleus. In such processes the
nucleus acts as discreet unit. Coherent processes can be easily identified
by the transverse momentum spectrum of the final state particle system.
They manifest themselves as diffractive peaks with large values
of the slope parameters determined
by the size of the nucleus: $dN/dP^2_T \simeq exp(-bP^2_T)$, where
$b\simeq 10~A^{2/3}$~GeV$^{-2}$. Owing to the difference in the absorption
of single-particle and multiparticle objects in nuclei, coherent processes
could serve as an effective tool for separation of resonance production against
non-resonant multiparticle background (see, e.g.~[35]).

The conditions for coherent reactions are destroyed by absorption processes in nuclei.
Thus, the coherent suppression of nonresonant background takes place:
\begin{equation}
\frac{\sigma_{coh}(res)}{\sigma_{coh}~(nonres.~BG)} >
\frac{\sigma_{noncoh}(res)}{\sigma_{noncoh}~(nonres.~BG)}.
\end{equation}
The separation of coherent reactions can be achieved by studying $dN/dP^2_T$
distributions for processes under investigation.

\begin{figure}[h]
\parbox[c]{6.5cm}{\caption{\tolerance=200\emergencystretch=10pt
$dN/dP^2_T$ distribution
for typical diffractive production reaction
$(p + N \rightarrow [\Sigma^oK^+] + N)$. The $P^2_T$-regions for coherent
reaction on carbon nuclei
$(P^2_T < 0.075 \div 0.1$~GeV$^2$) and for nonperipheral processes
$(P^2_T > 0.3$~GeV$^2$) are shown.}}
\hfill\parbox[c]{9cm}{\psfig{file=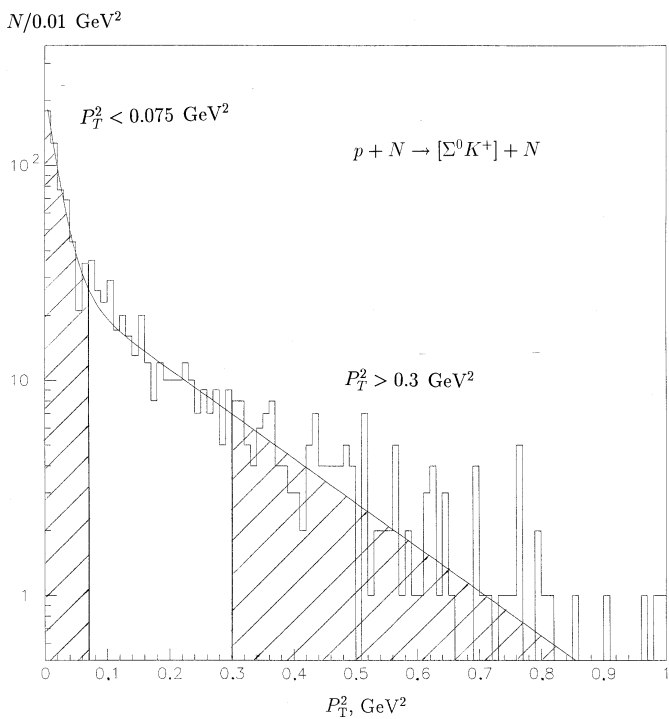,width=9cm}}
\end{figure}

As is seen from $dN/dP^2_T$ spectra in the SPHINX experiments, there are strong narrow forward
cones in these distributions with the slope $b\gtrsim 50$~GeV$^{-2}$,
which correspond to a coherent diffractive production on carbon nuclei
(see, for example, Fig.~2).
For identification of the coherently produced events we used ``soft''
transverse momentum cut $P^2_T <0.075 \div 0.1$~GeV$^2$.
With this cut noncoherent background in the event sample can be as large as
30$\div$40\%. It is possible to reduce this background with more stringent
$P^2_T$ cut at the cost of some decrease of the signal statistics.

\subsection{Processes with Large Transverse Momenta}

As was discussed above, coherent diffractive production reactions with small
transverse momenta seem quite promising for the search for exotic hadrons, but,
of course, they do not exhaust the existing opportunities. Certainly, these
searches can be also performed for all diffractive-type processes (e.g. without
coherent cuts for small $P^2_T$). And of special interest
is the study of nonperipheral production reactions which can be the
most effective way to seek for certain exotic states,
especially those that are formed at short ranges and are characterized
by broad enough transverse momentum distributions. In this case, the
most favorable conditions for identifying exotic hadrons are achieved at
higher transverse momenta ($P^2_T>0.3-0.5$~GeV$^2$), where
the background from peripheral processes is strongly suppressed. For instance,
the unusually narrow meson states $X(1740) \rightarrow \eta \eta$~[36] and
$X(1910) \rightarrow \eta \eta'$~[37] were observed in studying
the charge-exchange reactions $\pi^- + p \rightarrow [\eta \eta] + \Delta^o$ and
$\pi^- + p \rightarrow [\eta \eta'] + n$ after the selection of events with
$P^2_T \geq 0.3$~GeV. These anomalous states are good candidates for cryptoexotic
mesons. The rescattering mechanism involving multipomeron exchange in
the final state (a gluon-rich process) may explain X(1740) and X(1910)
nonperipheral production~[38].

In the very high primary energy region which, for example, corresponds
to the search for exotic states with heavy quarks, diffractive
production reactions with rescattering can be used, instead of the
charge-exchange processes with rescattering (see the diagram in Fig.~3). The cross
sections of these diffractive processes also do not die out with
energy rise.  The nonperipheral $P^2_T$ region for these processes is shown
in Fig.~2.

\begin{figure}[htb]
\centerline{\psfig{file=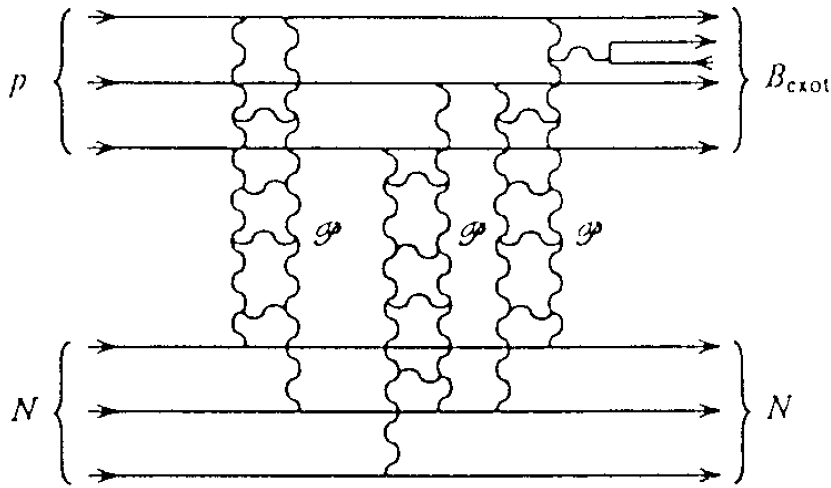,width=11cm,clip=}}
\caption{Diagram for diffractive~-- type reaction with multiple Pomeron
rescattering (these processes might be significant at $P^2_T \gtrsim 0.3$~GeV$^2$).}
\end{figure}

\subsection{Electromagnetic Mechanism}

The search for exotic hadrons with hidden strangeness can be also carried
out in another type of hadron production processes caused by
electromagnetic interactions. The example of such process is the formation
reaction with s-channel resonance photoproduction of strange particles
\begin{equation}
\gamma + N \rightarrow |qqqs \bar s > \rightarrow YK
\end{equation}
(see diagram in Fig.~4a). It is possible in principle to study
the s-channel resonance production by detailed energetic scanning of the cross sections
and angular distributions for (7) and by performing the subsequent
partial-wave analysis. As is seen from Fig.~4a and from VDM (with its significant
coupling of photon with $s \bar s$ pair through $\phi$-meson), reaction (7)
can provide a natural way to embed the $s \bar s$ quark pair into
intermediate resonance state and to produce the exotic baryon with
hidden strangeness.

\begin{figure}[htb]
\centerline{\psfig{file=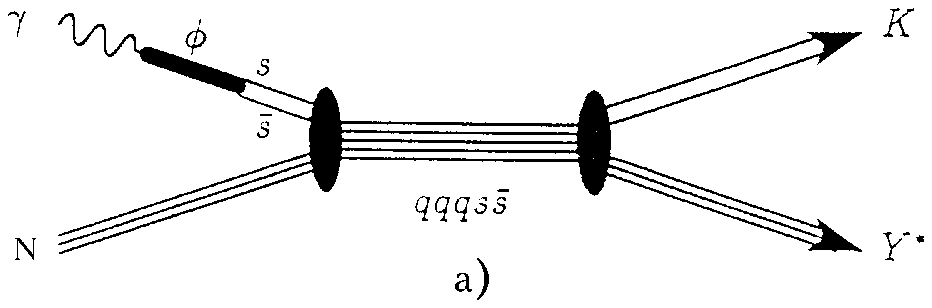,width=10cm}}


\centerline{\psfig{file=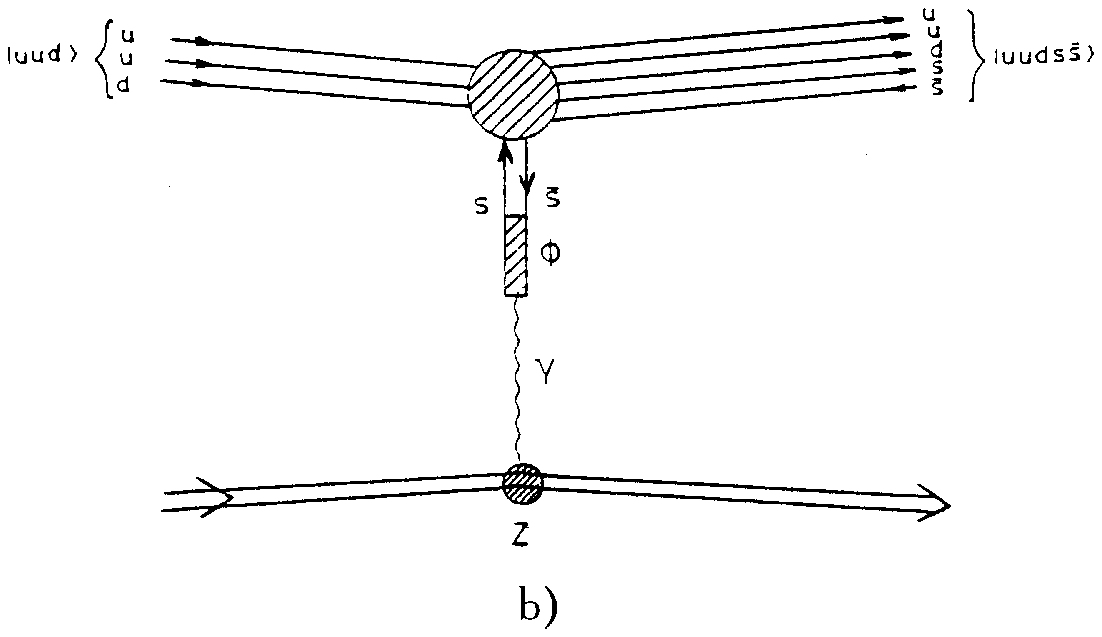,width=10cm}}
\caption{\small Electromagnetic~~mechanisms~~for~~production of exotic~~baryons
with~~~hidden~~~strangeness:\protect\linebreak a) formation reaction with $s$-channel resonance
photoproduction; b) Coulomb production reaction $h + (Z,A) \to a + (Z,A)$.}
\end{figure}

The existing data on reactions (7) are rather poor and insufficient for such
systematical studies. But one can hope that  good data would be produced
in the near future in the experiments on strong current electron
accelerators CEBAF and ELSA (see, for example,~[39]).

Electromagnetic production  of exotic hadrons can be searched for not
only in the resonance photoproduction reactions but in the collisions
of the primary hadrons with virtual photons of the Coulomb field of the target
nuclei~[40-44], e.g. in the Primakoff production reactions
\begin{equation}
h + Z \rightarrow a + Z
\end{equation}
(see diagram in Fig.~4b). The Coulomb production mechanism plays
the leading role in the region of very small transfer momenta, where it
dominates over the strong interaction process~[40-42]. The total cross section
of the Coulomb reaction (8) is
\begin{equation}
\sigma[h + Z \rightarrow a + Z]|_{Coulomb} \simeq \sigma_o~
\frac{2J_a+1}{2J_h +1}~\Gamma(a \rightarrow h + \gamma).
\end{equation}
The value $\sigma_o$ is obtained in the QED calculations. In the narrow
width approximation for the resonance a $\sigma_o$ has the form
\begin{equation}
\sigma_o = 8 \pi \alpha Z^2 \Bigl[ \frac{M_a}{M^2_a - m^2_h}\Bigr]^3
\int \limits_{q^2_{min}}^{q^2} \frac{[q^2 - q^2_{min}]}{q^4}|F_z(q^2)|^2 dq^2
\end{equation}
Here Z is the charge of nucleus; $\alpha = 1/137$ is the narrow
structure constant; $\Gamma(a \rightarrow h + \gamma)$ is the radiative
width of $a;~~J_a,~~J_h$ and $M_a,~~m_h$ are the spins and masses
of $a$ and $h$ particles; $F_z(q^2)$ is the electromagnetic formfactor
of nucleus; $q^2_{min}=\bigl[ [M^2_a - m^2_h]/2E_h \bigr]^2$ is the minimum square
momentum transfer $q^2$; $E_h$ is the primary hadron energy.
In the high energy region $q^2_{min}$ is very
small and $q^2 = P^2_T + q^2_{min} \simeq P^2_T$.

It must be borne in mind that in the Coulomb production
reactions with primary protons one studies the same processes as in ordinary
photoproduction reactions. But the Coulomb production in the experiments with unstable
primary particles (pions, kaons, hyperons) opens the unique possibility to study
the photoproduction reactions with these unstable ``targets''.

\section{THE EXPERIMENTS WITH THE SPHINX SETUP}

The search for exotic baryons with hidden strangeness was performed in the
experiments on the proton beam of the 70~GeV IHEP accelerator with the SPHINX
spectrometer. The SPHINX setup includes the following basic components:

1. A wide aperture magnetic spectrometer with proportional wire chambers,
drift chambers, drift tubes, scintillator hodoscopes.

2. Multichannel $\gamma$-spectrometer with lead-glass Cherenkov total
absorption counters.

3. A system of gas Cherenkov detectors for  identification of secondary
charged particles (including a RICH detector with photomatrix equipped with
736 small phototubes; this is the first counter of this type used in the
experiments).

4. Trigger electronics, data acquisition system and on-line control system.

The SPHINX spectrometer works in the proton beam with energy
$E_p$=70~GeV and intensity $I \approx (2\div 3)\cdot 10^6$~p/cycle.
The measurements were performed with a polyethylene target to optimize the
acceptance, sensitivity and secondary photon losses.

The first version of the SPHINX setup was described in~[12]. The next version of
this setup after partial modification (with a new  $\gamma$-spectrometer
and with better conditions for $\Lambda$ and $\Sigma^o$ identification)
was discussed in~[21].

To separate different exclusive reactions, a complete kinematical
reconstruction of events was performed by taking into account information
from the tracking detectors, from the magnetic spectrometer, from the
$\gamma$-spectrometer, and from all Cherenkov counters of the SPHINX
setup. At the final stage of this reconstruction procedure, the
reactions under study were identified by examining the effective-mass spectra for
subsystems of secondary particles.

Several photon-induced diffractive production processes were studied in the
experiments of the SPHINX Collaboration~[12-25;27;28]:
\begin{eqnarray}
p + N \rightarrow &\!& [p K^+ K^-] + N, \\
\rightarrow &\!& [p \phi]+ N \\
&\!&~~~\protect \raisebox {1.18ex}{$ \lfloor$} \! \! \! \nonumber
\rightarrow K^+ K^- \nonumber  \\
 \rightarrow &\!& [\Lambda(1520)K^+] + N,
\label{eq:4}\\
&\!&~\protect \raisebox {1.11ex}{$ \lfloor$} \! \! \! \nonumber
\rightarrow K^- p  \nonumber  \\
 \rightarrow &\!& [\Sigma^*(1385)^oK^+] + N,
\label{eq:4}\\
&\!&~\protect \raisebox {1.11ex}{$ \lfloor$} \! \! \! \nonumber
\rightarrow \Lambda \pi^o  \nonumber \\
 \rightarrow &\!& [\Sigma^*(1385)^oK^+] + N + (neutrals),
\label{eq:4}\\
&\!&~\protect \raisebox {1.11ex}{$ \lfloor$} \! \! \! \nonumber
\rightarrow \Lambda \pi^o  \nonumber  \\
\rightarrow &\!& [\Sigma^oK^+] + N,\label{eq:4}\\
&\!&~\protect \raisebox {1.11ex}{$ \lfloor$} \! \! \! \nonumber
\rightarrow \Lambda \gamma  \nonumber \\
\rightarrow &\!& [\Sigma^+ \hspace*{7mm}  K^o] + N \\
&\!&\hspace{2mm}^|\hspace{-2mm} \rightarrow p \pi^0
 \hspace{2mm}^|\hspace{-2mm} \rightarrow \pi^+ \pi^- \nonumber \\
\rightarrow &\!& [p \eta] + N,
\label{eq:4}\\
&\!&~~~\protect \raisebox {1.14ex}{$ \lfloor$} \! \! \! \nonumber
\rightarrow \pi^+ \pi^- \pi^o  \nonumber \\
\rightarrow &\!& [p \eta'] + N,
\label{eq:4}\\
&\!&~~~\protect \raisebox {1.14ex}{$ \lfloor$} \! \! \! \nonumber
\rightarrow \pi^+ \pi^- \eta \rightarrow \pi^+ \pi^- 2\gamma  \nonumber \\
\rightarrow &\!&[ p \omega] + N,
\label{eq:4}\\
&\!&~~~\protect \raisebox {1.18ex}{$ \lfloor$} \! \! \! \nonumber
\rightarrow \pi^+ \pi^- \pi^o \\ 
\rightarrow &\!& [p \pi^+ \pi^-] + N,
\label{eq:4}\\
\rightarrow &\!&[\Delta^{++} \pi^-]+ N,
\label{eq:4}\\
&\!&~\protect \raisebox {1.11ex}{$ \lfloor$} \! \! \! \nonumber
\rightarrow p \pi^+ \\ \nonumber
\end{eqnarray}
and several other processes. Here N is nucleon or C nucleus for the coherent
processes.
The separation of coherent diffractive processes is obtained by studying
their $dN/dP^2_T$ distributions, as is shown in Fig.~2.

\section{PREVIOUS~DATA~ON~THE~COHERENT
DIFFRACTIVE REACTIONS
\boldmath {$p+C  \rightarrow [\Sigma^o K^+]+C$}
AND \boldmath{$p+C \rightarrow [\Sigma^*(1385)^o K^+]+C$}}

One of the major results obtained with the SPHINX setup
was the study of $\Sigma^o K^+$ system produced in  diffractive process (16).

These data
were obtained in two different runs on the SPHINX facility:

a) the first run with the old version of this setup (``old run'',~[12,17-20]);

b) the second run with partially upgraded SPHINX apparatus
(``new run''~[21-23]). As a result of this upgrading the detection efficiency
and purity of $\Lambda$ and $\Sigma^o$ events were significantly increased.

The main results of these measurements can be summarized as follows:

1. Old~[16-20] and new [21-23] data from  coherent diffractive
reaction (16) were obtained under different experimental
conditions, with a significantly modified apparatus, with
different background and systematics. Nevertheless, the $\Sigma^o K^+$ invariant
mass spectra from both runs are in a good agreement which makes them more reliable.

2. The combined mass spectrum $M(\Sigma^oK^+)$ for  coherent reaction (16)
from the old and new data (with $P^2_T < 0.1$~GeV$^2$) is presented in Fig.5.
This spectrum is dominated by the X(2000) peak with parameters in Table~1.

3. There is also some near threshold structure in this $M(\Sigma^oK^+)$
spectrum in the region of $\sim 1800$~MeV (see Fig.5 and Table~1). Such a shape of
the $\Sigma^oK^+$ mass spectrum (with an additional structure near the
threshold) proves that the X(2000) peak cannot be explained by a
non-resonant Deck-type diffractive singularity. Therefore, most likely this
peak has a resonant nature.

\begin{figure}[htb]
\centerline{\psfig{file=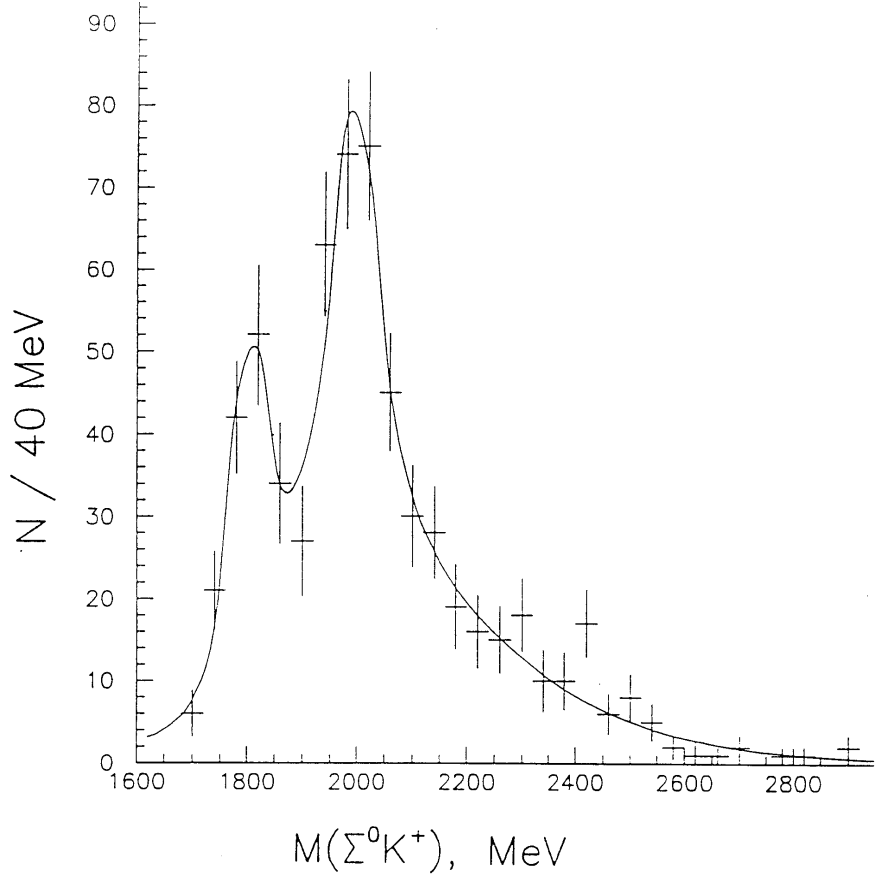,width=12cm}}
\caption{Combined mass spectrum $M(\Sigma^oK^+)$ for coherent diffractive
reaction (16) in old and new runs on the SPHINX setup $(P^2_T < 0.1$~GeV$^2$).
The parameters of $X(2000)$ peak in this spectrum are: $M = 1997 \pm 7$~MeV;
$\Gamma = 91 \pm 17$~MeV.}
\end{figure}

\newpage
\vspace*{-2.0cm}
\begin{table}[H]
\small
\caption{\small The main results of the previous SPHINX data for coherent
diffractive reactions $p+C \rightarrow [\Sigma^o K^+]+C$ and
$p+C \rightarrow [\Sigma^*(1385)^oK^+]+C$}
\vspace*{-0.2cm}
\begin{center}
\begin{tabular}{l p{13cm}}\hline
\label{tab-1}
1. & Coherent diffractive production reaction $p+C \rightarrow [\Sigma^o K^+]+C$
with  coherent cut $P^2_T < 0.1$~GeV$^2$ was studied in the old and new runs.
The combined mass spectrum $M(\Sigma^oK^+)$ is presented on Fig.~5. This spectrum
is dominated by X(2000) state with parameters\\[-0.5cm]
& \begin{displaymath}
\left.  \begin{array}{ccc}
M~ =~ 1997 \pm 7~{\rm{MeV}},\\
\Gamma~= ~~91 \pm 17~{\rm{MeV}},\\
\multicolumn{3}{l}{{\rm{statistical~significance~of~the~peak~is~7~s.d.}}}\\
\end{array} \right \}.
\end{displaymath} \\[-0.7cm] \hline
2. & There are also some near threshold structure X(1810), which is produced
only in the region of very small $P^2_T$  ($ \lesssim 0.01 \div 0.02$~GeV$^2$).
The parameters of this peak are
$$ M=1812 \pm ~7~{\rm{MeV}}, $$
$$ \Gamma = ~~56 \pm 16~{\rm{MeV}}.$$
\\ [-0.5cm]\hline
3. & Coherent diffractive production reaction $p+C \rightarrow [\Sigma^*(1385)K^+]+C$
with tight coherent cut $P^2_T<0.02$~GeV$^2$
was studied in the
old run~([12;17-20]).
The mass spectra of $M[\Sigma^*(1385)^oK^+]$ are in Fig.~6. The peak was
observed in these spectra with  average value of parameters\\ [-0.5cm]
& \begin{displaymath}
\left.  \begin{array}{ccc}
M~ =~ 2052 \pm 6~{\rm{MeV}}\\
\Gamma~= ~~35^{+22}_{-35}~{\rm{MeV}}\\
\multicolumn{3}{l}{{\rm{(with~ the~ account~ of~ the~ apparatus~ mass~ resolution);}}}\\
\multicolumn{3}{l}{{\rm{statistical~ C.L.~ of~ the~ peak~  \geq 5~ s.~d.}}}\\
\end{array} \right \}.
\end{displaymath} \\[-0.7cm] \hline
4. & The data of (14) and (16) were analyzed together with the data from (21) and (22)
to obtain the branching ratios of
different decay channels (with strange particles and without strange
particle in final state). The lower limits of the ratios
were obtained from this comparative analysis (with 95\% C.L.): \\[-0.2cm]
&
$$
R_1= \frac{BR\{X(2050)^+ \rightarrow [\Sigma^*(1385)K]^+\}}
{BR\{X(2050)^+ \rightarrow [\Delta(1232) \pi]^+ \}} >1.7
$$
$$
R_2= \frac{BR\{X(2050)^+ \rightarrow [\Sigma^*(1385)K]^+\}}
{BR\{X(2050)^+ \rightarrow p \pi^+ \pi^-\}} >2.6
$$
$$
R'_2= \frac{BR\{X(2050)^+ \rightarrow \Sigma^*(1385)^o K^+\}}
{BR\{X(2050)^+ \rightarrow p \pi^+ \pi^-\}} >0.86
$$
$$
R_3 = \frac{BR\{X(2000)^+ \rightarrow [\Sigma K]^+\}}
{BR\{X(2000)^+ \rightarrow [\Delta (1232)\pi]^+\}} >0.83
$$
$$
R_4= \frac{BR\{X(2000)^+ \rightarrow [\Sigma K]^+\}}
{BR\{X(2000)^+ \rightarrow p \pi^+ \pi^-\}} >7.8
$$
$$
R'_4 = \frac{BR\{X(2000)^+ \rightarrow \Sigma^o K^+\}}
{BR\{X(2000)^+ \rightarrow p \pi^+\pi^-\}} >2.6.
$$
\\[-0.5cm] \hline
\end{tabular}
\end{center}
\end{table}

A strong influence of $P^2_T$ cut for the production of this X(1810) state
was established: this structure is produced only at very small $P^2_T$
$(\lesssim 0.01 \div 0.02~$GeV$^2$)~--- see below.

We have also some data in studying  the $\Sigma^*(1385)^oK^+$ system
in reaction (14), which were obtained only in the old run (data on this
reaction from the new run are now in process of analysis).
Coherent events of (14) were singled out in the analysis of $dN/dP^2_T$
distribution as a strong forward peak with the slope $b \gtrsim 50$~GeV$^{-2}$.
In order to reduce the noncoherent background and to obtain
the $\Sigma(1385)^oK^+$ mass spectrum for the ``pure''
coherent production reaction on carbon nuclei a ``tight''
requirement $P^2_T<0.02$~GeV$^2$ was imposed and the mass spectra of
$\Sigma(1385)^oK^+$ for the coherent events of (14) were obtained (see, for
example, Fig.~6).
In these spectra some very narrow structure X(2050) was observed.
The fits of the spectra with Breit-Wigner peaks and polynomial
smooth background were carried out, and the average values for the main
parameters of $X(2050)$ structure are presented in Table~1. Certainly, one
needs further confirmation of the existence of X(2050) in
the new data with increased statistics. Up to now it is impossible
also to exclude completely the feasibility for X(2000) and X(2050)
to be in fact two different decay modes of the same state.

\begin{figure}[htb]
\parbox[c]{7.5cm}{\psfig{file=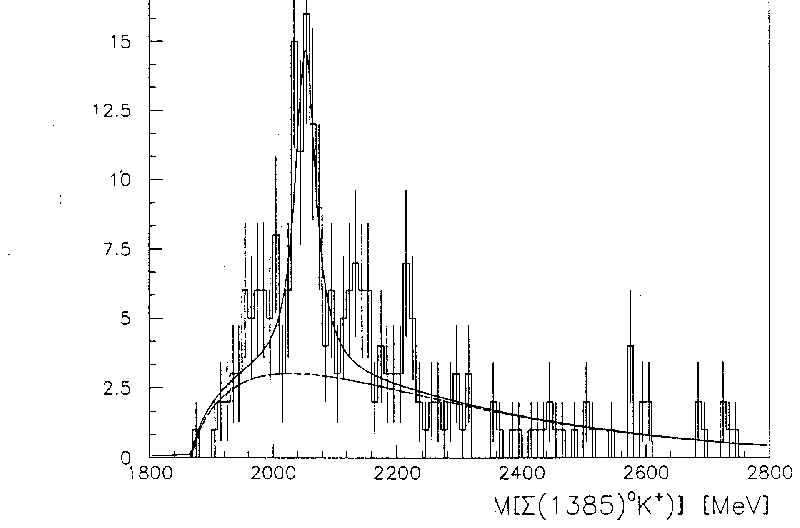,width=7.5cm}}
\hfill\parbox[c]{7.5cm}{\psfig{file=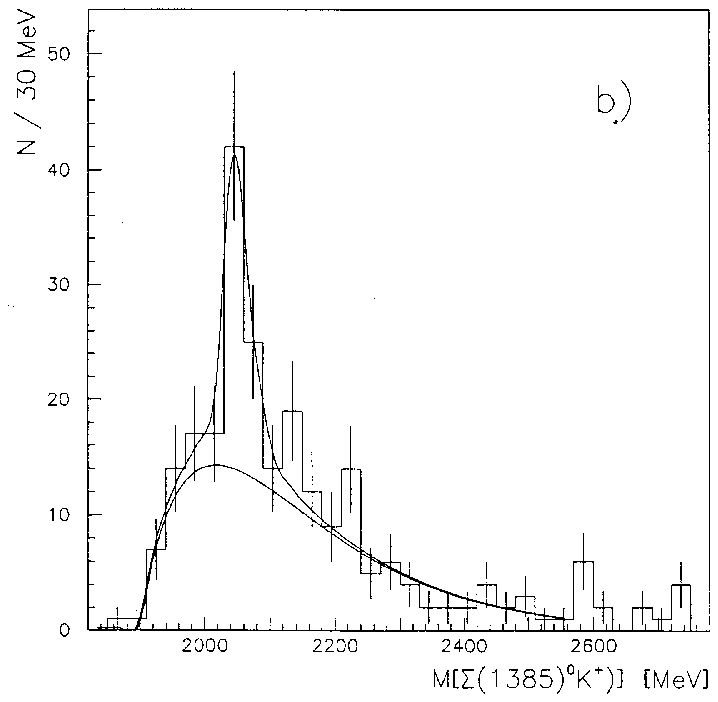,width=7.5cm}}

\caption{Invariant mass spectra M $[\Sigma^*(1385)^oK^+]$ in the coherent
reaction (14) at tight transverse-momentum cut $P^2_T<0.02$~GeV$^2$ for
various bin widths: a) $\Delta M$=10~MeV; b)~$\Delta M$=30~MeV.
The spectra are fitted to the sum of a smooth polynomial background
and X(2050) Breit-Wigner peak. The parameters of X(2050) peak are:
a)~$M=2053 \pm 4$~MeV, $\Gamma=40 \pm 15$~MeV;
b)~$M=2053\pm5$~MeV, $\Gamma=35\pm16$~MeV.}
\end{figure}

In  studying  coherent reactions (21)
$p + C  \rightarrow  [p \pi^+\pi^-] + C$  and (22) $p+C \rightarrow [\Delta(1232)^{++}
\pi^-]+C$
under the same kinematical conditions as of processes (14) and (16)
a search for other decays
channels of the X(2000) and X(2050)
baryons was performed~[18,19]. No peaks in 2~GeV mass
range were observed in the mass spectra of $p\pi^+\pi^-$ and $\Delta(1232)^{++}\pi^-$
systems produced in reactions (21) and (22), respectively. Lower limits on the
corresponding decay branching ratios
R (see (5)) were obtained from this comparative analysis:
\begin{equation}
R[X(2000);~X(2050)] \gtrsim 1 \div 10 \qquad {\rm{(95\%~~C.L.)}}
\end{equation}
(more details are in Table~1).

The isotopic relations between the decay amplitudes of  $I = 1/2$ particles
were assumed in these calculations (the $X(2000)$ and X(2050)
states belong to  isodoublets
since they are  produced in a diffractive dissociation of  proton).
In accordance with these relations
\begin{equation}
BR[X^+_{I=1/2} \rightarrow \Sigma^0K^+] = \frac{1}{3} BR[X^+_{I=1/2}
\rightarrow (\Sigma K)^+]
\end{equation}
\begin{equation}
BR[X^+_{I=1/2} \rightarrow \Delta^{++}\pi^-] = \frac{1}{2}BR[X^+_{I=1/2}
\rightarrow (\Delta \pi)^+]
\end{equation}

The ratios $R_1 - R_4$ of the widths of the $X(2000)$ and X(2050) decays into strange and
nonstrange particles are much larger than those for ordinary $(qqq)$-isobars~[18,33].

A narrow width of the $X(2000)$ and X(2050) baryon states as well as anomalously large
branching ratios for their decay channels with strange particle emission (large
values of $R$) are the reasons to consider these states as  serious
candidates for cryptoexotic baryons with a hidden strangeness $|uuds \bar s>$.

\section{NEW ANALYSIS OF THE DATA
FOR~REACTION~\boldmath{$p + N \rightarrow [\Sigma^oK^+] + N$}}

In what follows we present the results of a new analysis of the data
obtained in the run with partially upgraded SPHINX spectrometer where conditions
for $\Lambda$ and $\Sigma^o$ separation were greatly improved as compared with
an old version of this setup. The key element of the new analysis lays in a
detailed study of the $\Sigma^o \rightarrow \Lambda + \gamma$ decay separation,
which makes it possible to reach the reliable identification of this
decay and
reaction (16) with the increased efficiency in comparison with the previous
analysis of~[21].

In this new analysis the data for (16) were studied with different criteria
for $\Sigma^o \rightarrow \Lambda + \gamma$ separation (with larger efficiency
and larger background or with the reduced background at the price of lower
efficiency). We will designate these different criteria for photon separation
as soft, intermediate and strong photon cuts (the details will be presented
in~[45]). Reaction (16) was studied in different kinematical regions. It was
found that improved background conditions were important  for
the investigation of the region of small mass $M(\Sigma^oK^+)$ and
very small transverse momenta.

The effective mass spectra $M(\Sigma^o K^+)$ in $p + N \rightarrow [\Sigma^oK^+] + N$
 for all $P^2_T$ are
presented in Fig.~7. The peak of
$X(2000)$ baryon state with $M=1986\pm6$~MeV and $\Gamma=98\pm20$~MeV is seen
very clearly in these spectra with a very good statistical significance.
Thus, the reaction
\begin{eqnarray}
p + N \rightarrow &\!& X(2000) + N
\label{eq:4},\\
&\!&~\protect \raisebox {1.11ex}{$ \lfloor$} \! \! \!
\rightarrow \Sigma^o K^+ \nonumber
\end{eqnarray}
is well separated in the SPHINX data. The cross section
for $X(2000)$ production in (26) is:
\begin{equation}
\sigma[p+N \rightarrow X(2000)+N] \cdot BR[X(20000) \rightarrow \Sigma^o K^+]=
95\pm20~{\rm{nb/nucleon}}
\end{equation}
(with respect to one nucleon under the assumption of $\sigma \propto A^{2/3}$,
e.g. for the effective number of nucleons in carbon nucleus
equal to 5.24). The parameters of X(2000) are not sensitive to the different photon cuts,
as  is seen from Table~2.
The $dN/dP^2_T$ distribution for reaction (26) is shown in Fig.~8. From
this distribution the coherent diffractive production reaction on
carbon nuclei is identified as a diffraction peak with the slope
$b\simeq 63\pm10$~GeV$^{-2}$. The cross section for coherent
reaction is determined as
$$
\sigma[p+C \rightarrow X(2000)^+ + C]_{\rm{Coherent}} \cdot
BR[X(2000)^+ \rightarrow \Sigma^o K^+]=$$
\begin{equation}
= 260\pm60~{\rm{nb/C~nucleus}}.
\end{equation}


\newpage
\begin{center}
\vspace*{-3.5cm}
\end{center}

\small
\begin{table}[htb]
\caption{Data on $M(\Sigma^0K^+)$ in reaction $p + N \rightarrow [\Sigma^0K^+] + N$,
$\Sigma^0 \rightarrow \Lambda \gamma$ with different photon cuts (for all $P^2_T$)}
\begin{center}
\vspace*{-0.3cm}
\begin{tabular}{ll|c|c|c}\hline
\multicolumn{2}{c|}{Photon cut} & Soft & Intermediate & Strong \\ \hline
\multicolumn{2}{c|}{$N$ events in $X(2000)$ peak} & $430 \pm 89$ & $301 \pm 71$ & $190 \pm 47$ \\ \hline
\multicolumn{2}{c|}{Correction factor for photon} & 1.0 & 1.4 & 2.25 \\
\multicolumn{2}{c|}{efficiency} & & & \\ \hline \hline
\multicolumn{2}{c|}{Parameters of $X(2000)$} & & & \\ \hline
$M$ (MeV) & weighted spectrum & $1986 \pm 6$ & $1991 \pm 8$ & $1988 \pm 6$ \\ \cline{2-5}
& measured spectrum & $1988 \pm 5$ & $ 1994 \pm 7$ & $1990 \pm 6$ \\ \hline
$\Gamma$ (MeV) & weighted spectrum & $98 \pm 20$ & $96 \pm 26$ & $68 \pm 21$ \\
 \cline{2-5}
& measured spectrum  & $84 \pm 20$ & $94 \pm 21$ & $68 \pm 20$ \\
 \hline \hline
\multicolumn{2}{l|}{$\sigma[p + N \rightarrow X(2000) + N] \cdot$} &
$100 \pm 19$ & $93 \pm 25$ & $91 \pm 21$ \\
\multicolumn{2}{l|}{$\cdot BR[X(2000) \rightarrow \Sigma^0K^+]$} & & & \\
\multicolumn{2}{l|}{(nb/nucleon)} & & & \\ \hline
& \multicolumn{1}{|c|}{$<M>$ MeV} & \multicolumn{3}{|c}{$1989 \pm 6$} \\ \cline{2-5}
& \multicolumn{1}{|c|}{$<\Gamma>$ MeV} & \multicolumn{3}{|c}{$91 \pm 20$} \\ \cline{2-5}
Average& \multicolumn{1}{|c|}{$< \sigma[p + N \rightarrow X(2000) + N]> \cdot$}
& \multicolumn{3}{|c}{$95 \pm 20$ (statist.) $\pm 20$ (system.)} \\
values& \multicolumn{1}{|c|}{$\cdot BR[X(2000) \rightarrow \Sigma^0K^+]$}
& \multicolumn{3}{|c}{} \\
& \multicolumn{1}{|c|}{nb/nucleon}
& \multicolumn{3}{|c}{} \\ \cline{2-5}
& \multicolumn{1}{|c|}{$<\sigma[p + C \rightarrow X(2000) + C]> \cdot$}
& \multicolumn{3}{|c}{$285\pm60$ (statist.) $\pm 60$ (system.)} \\
& \multicolumn{1}{|c|}{$\cdot BR[X{2000} \rightarrow \Sigma^0K^+]$}
& \multicolumn{3}{|c}{} \\
& \multicolumn{1}{|c|}{nb/C nucleus}
& \multicolumn{3}{|c}{} \\ \hline
\end{tabular}
\end{center}
\end{table}

\normalsize

\vspace*{-0.5cm}

\begin{figure}[H]
\centerline{\psfig{file=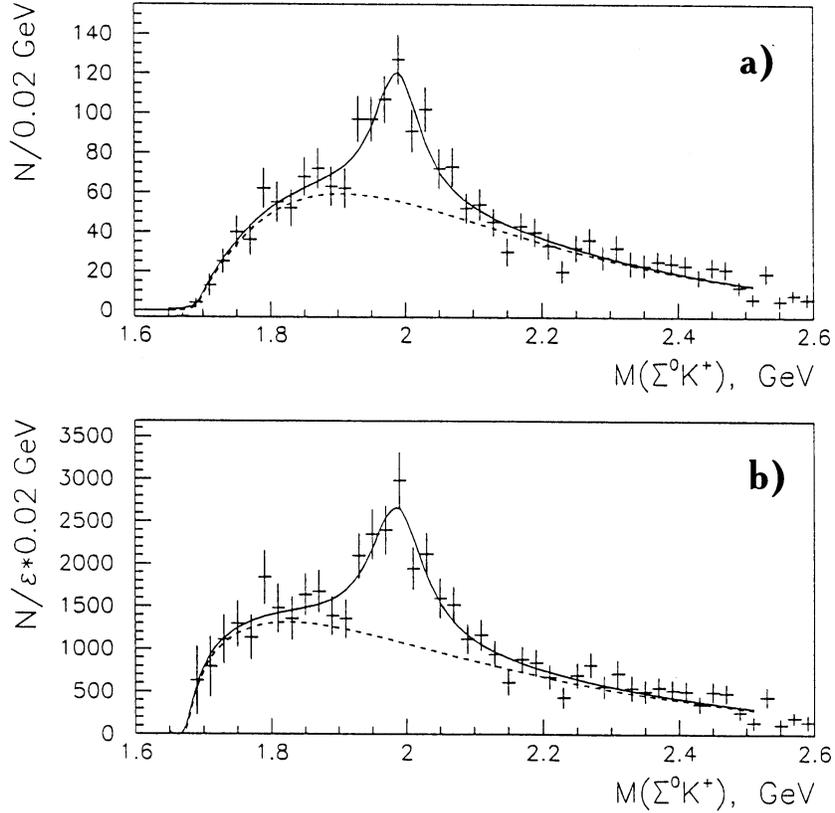,width=11cm,clip=}}
\vspace*{-0.3cm}\caption{Invariant mass spectra $M(\Sigma^oK^+)$ in the diffractive reaction
$p + N \rightarrow [\Sigma^oK^+] + N$ for all $P^2_T$ (with soft photon cut):
a) measured mass spectrum; b) the same mass spectrum weighted with the
efficiency of the setup. Parameters of $X(2000)$ peak are in Table~2.}
\end{figure}

\newpage

\begin{figure}[htb]
\vspace*{0.5cm}
\centerline{\psfig{file=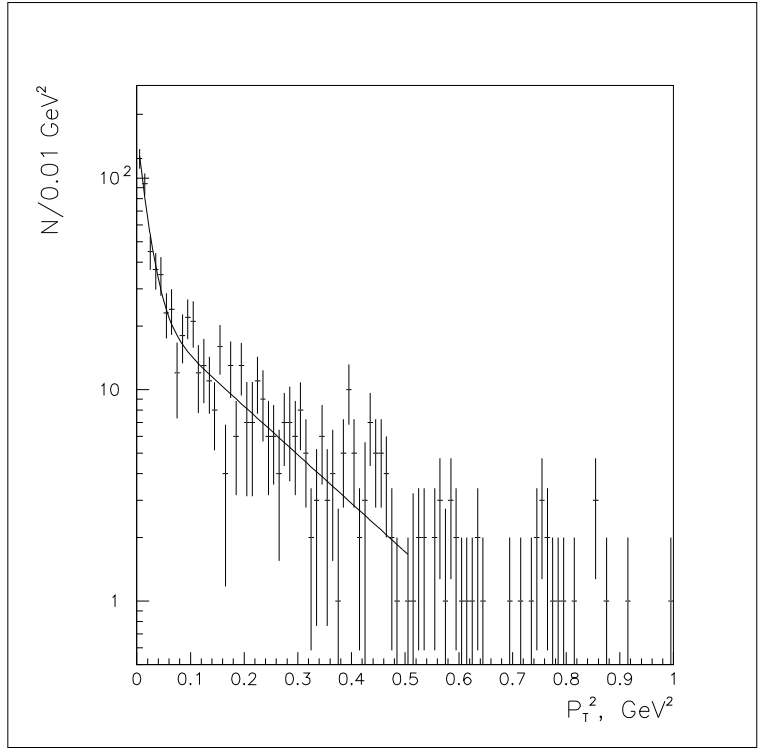,clip=,width=9cm}}
\caption{$dN/dP^2_T$ distribution for the diffractive production reaction
$p + N \rightarrow X(2000) + N$. The distribution is fitted in the form
$dN/dP^2_T = a_1$ $exp(-b_1P^2_T) + a_2exp(-b_2P^2_T)$ with the slope parameters
$b_1 =63 \pm 10$~GeV$^{-2}$; $b_2=5.8\pm 0.6$~GeV$^{-2}$.}
\end{figure}
\vspace*{-0.2cm}

We must bear in mind that
it is more convenient to use other relations for cross sections:
\vspace*{-0.1cm}
\begin{equation}
\sigma[p+N \rightarrow X(2000)^+ + N] \cdot BR[X(2000)^+ \rightarrow
(\Sigma K)^+]=285\pm~60~{\rm{nb/nucleon}},
\vspace*{-0.1cm}
\end{equation}
\vspace*{-0.4cm}
\begin{equation}
\sigma[p+C \rightarrow X(2000)^+ + C] \cdot BR[X(2000)^+ \rightarrow
(\Sigma K)^+]=780 \pm 180~{\rm{nb/nucleus}},
\vspace*{-0.1cm}
\end{equation}
which were obtained from (27) and (28) using branching ratio (24).

The errors in the cross sections of (27)-(30) are statistical only. Additional systematic
errors are $\simeq \pm 20\%$ due to uncertainties in the cuts, in Monte
Carlo efficiency calculations and in the absolute normalization.

In the mass spectra $M(\Sigma^oK^+)$ in Fig.~7 there is only a slight indication
for $X(1810)$ structure which was observed earlier in the study of coherent
reaction (16)~--- see Fig.~5 and~[21]. This difference is caused by a large
background in this region for the events in Fig.7 (all $P^2_T$, soft photon
cut). To clarify the situation in this new analysis we
investigated also the $M(\Sigma^oK^+)$ mass spectra for coherent
reaction (16), e.g. for $P^2_T <0.1$~GeV$^2$. In these mass spectra
not only the X(2000) peak is observed, but the X(1810) structure
as well. These spectra (see~[45]) are compatible with the data in Fig.~5.

The yield of the X(1810) as function of $P^2_T$ is shown in Fig.~9. From
this figure
it is clear that X(1810) is produced only
in a very small $P^2_T$ region ($P^2_T<0.01-0.02$~GeV$^2$). For
$P^2_T < 0.01$~GeV$^2$ the $M(\Sigma^o K^+)$ mass spectrum
demonstrates a very sharp X(1810) signal (see Fig.~10) with the
parameters of the peak
\vspace*{-0.1cm}
\begin{equation}
X(1810) \rightarrow \Sigma^oK^+
\left \{  \begin{array}{ccc}
M & = & 1807 \pm 7~{\rm{MeV}}\\
\Gamma& = & ~~62 \pm 19~{\rm{MeV}}
\end{array} \right.
\vspace*{-0.1cm}
\end{equation}
which is in a  good agreement with the previous data of Table~1.
The cross section for coherent X(1810) production is
\vspace*{-0.2cm}
$$
\sigma[p+C \rightarrow X(1810)^+ +C]|_{P^2_T<0.01~{\rm{GeV^2}}} \cdot
BR[X(1810)^+ \rightarrow \Sigma^o K^+] =$$
\vspace*{-0.1cm}
\begin{equation}
= 215 \pm 45~{\rm{nb/C~nucleus}}.
\vspace*{-0.1cm}
\end{equation}

\vspace*{-1.5cm}

\begin{figure}[htb]
\parbox[c]{6.5cm}{\psfig{file=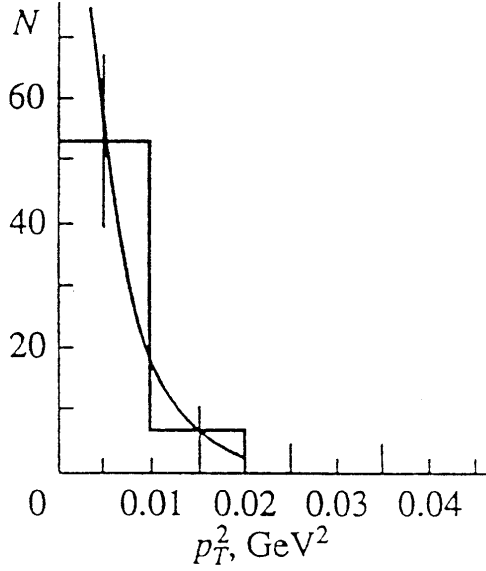,width=6.5cm}}
\hfill\parbox[c]{10cm}{\caption{The $P^2_T$ dependence for the $X(1810)$ structure production in the
coherent reaction $p + C \rightarrow X(1810) + C$.}}
\end{figure}

\vspace*{-0.5cm}

\begin{figure}[H]
\centerline{\psfig{file=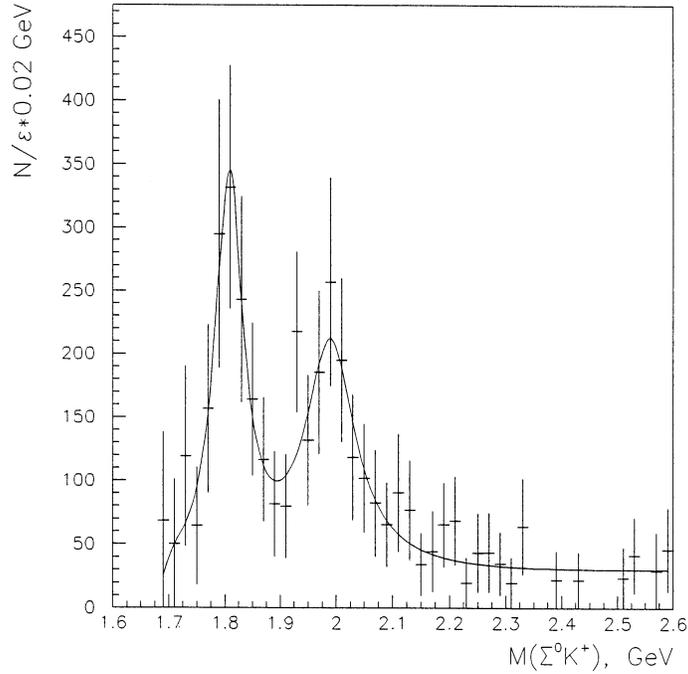,width=9cm,clip=}}
\caption{Invariant mass spectra $M(\Sigma^oK^+)$ in the coherent diffractive
production reaction $p + C \rightarrow [\Sigma^oK^+] + C$ in the region
of very small $P^2_T < 0.01$~GeV$^2$ (with strong photon cut)
weighted with the efficiency of the setup.}
\end{figure}

\newpage

The additional systematic error for this value is $\pm 30\%$. It increased
as compared with the same errors in (27)-(30) due to the uncertainty in the
evaluation of $P^2_T$ smearing in the region of $P^2_T<).01$~GeV$^2$, which
is more sensitive to $P^2_T$ resolution.

It is possible also to demonstrate the coherent diffractive X(2000)
production in the clearest way by using the ``restricted coherent region''
$0.02<P^2_T<0.1$~GeV$^2$ (see Fig.~11) where there is no influence of X(1810)
structure.
\begin{figure}[htb]
\centerline{\psfig{file=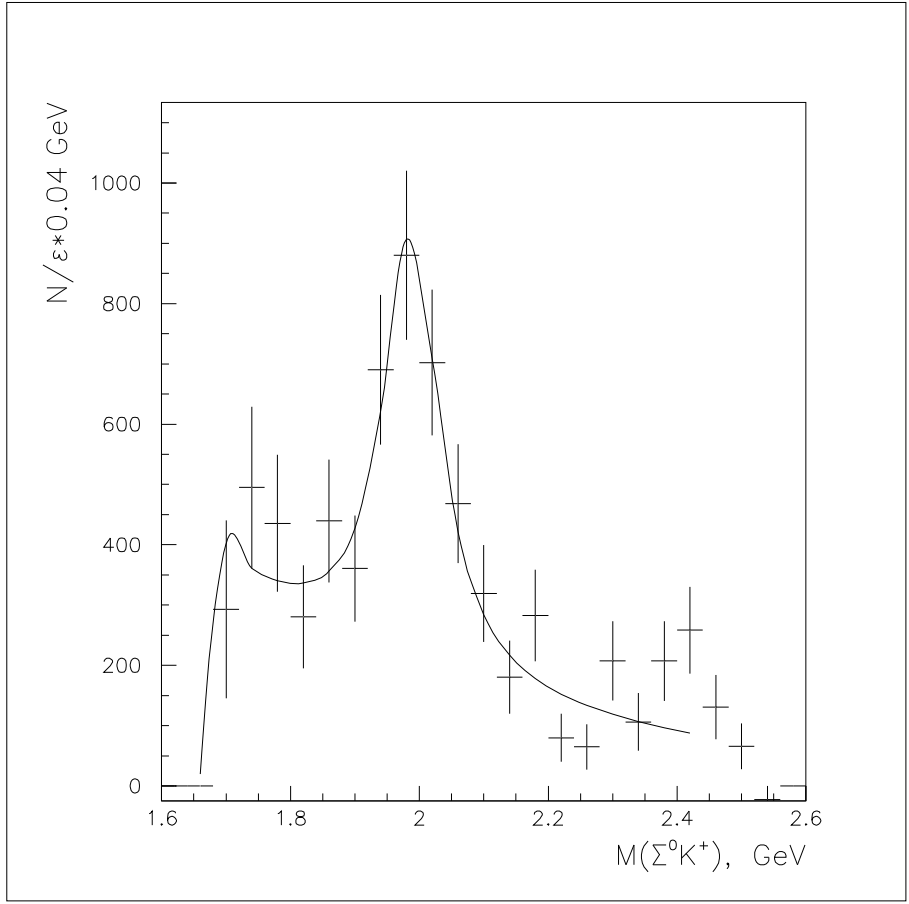,width=12cm,clip=}}
\caption{Weighted invariant mass spectrum $M(\Sigma^oK^+)$ for the reaction
$p + C \rightarrow [\Sigma^oK^+] + C$ in the ``restricted coherent region''
$0.02 < P^2_T < 0.1$~GeV$^2$ (with intermediate photon cut).}
\end{figure}

To explain the unusual properties of X(1810)
state in a very small $P^2_T$ region the hypothesis of the electromagnetic
production of this state in the Coulomb field of carbon nucleus was
proposed earlier [46]. It is possible to estimate the cross section
for the Coulomb X(1810) production from (9) and (10):
$$\sigma[p+C \rightarrow X(1810)^+ +C]|_{P^2_T<0.01~{\rm{GeV^2;~Coulomb}}} =$$
$$=(2J_x+1)\{\Gamma[X(1810)^+ \rightarrow p + \gamma][{\rm{MeV}}]\} \cdot
2.8 \cdot 10^{-30}~{\rm{cm^2/C~nucleus}} \geq$$
\begin{equation}
\geq 5.6 \cdot 10^{-30}~{\rm{cm^2}}\{\Gamma[X(1810)^+ \rightarrow p + \gamma]
[{\rm{MeV}}]\}
\end{equation}
($J_x \geq 1/2$ is the spin of X(1810)).

Let us compare this Coulomb hypothesis prediction with
the experimental value
\begin{equation}
\sigma[p+C \rightarrow X(1810)^+ + C]|_{P^2_T<0.01~{\rm{GeV^2}}} \gtrsim
645~{\rm{nb/C~nucleus}}.
\end{equation}
To obtain (34) we assumed in (16) that X(1810) is isodoublet, and then we
use from (24) the branching $BR(X^+ \rightarrow \Sigma^o K^+) \leqslant 1/3$
(here $\simeq$ means that $BR[X^+ \rightarrow (\Sigma K)^+] \simeq 1$,
i.e. this decay is dominating).

If the value of radiative width $\Gamma[X(1810) \rightarrow p+ \gamma]$ is
around 0.1-0.3~MeV and the branching $BR[X(1810)^+ \rightarrow
(\Sigma K)^+]$ is significant, then the experimental data for
cross section of the coherent X(1810) production (34) can be in agreement
with the Coulomb mechanism  prediction (33). It seems that such value of
radiative width is quite reasonable. For example, the radiative width for
$\Delta(1232)$ isobar is $\Gamma[\Delta(1232)^+ \rightarrow p + \gamma] \simeq$0.7~MeV.
The value of radiative width depends on the amplitude of this process A and on
kinematical factor: $\Gamma=|A|^2 \cdot (P_\gamma)^{2l+1}$ ($P_{\gamma}$ is
the momentum of photon in the rest frame of the decay baryon
and $l$ is orbital momentum). For $X(1810) \rightarrow p + \gamma$
decay the kinematical factor may be by an order of magnitude larger than for
$\Delta(1232)^+ \rightarrow p + \gamma$ because of the large mass of
X(1810) baryon. Certainly, the predictions for amplitude A are quite
speculative. But if, for example, X(1810) is the state with hidden
strangeness $|qqqs \bar s>$, then the amplitude A might be not very small
due to a possible VDM decay mechanism $(qqqs \bar s) \rightarrow (qqq) + \phi_{\rm{virt}}
\rightarrow (qqq) + \gamma$. Thus it seems that the experimental data for
the coherent production of X(1810) (34) is not in contradiction
with the Coulomb production hypothesis.

It is possible  that the candidate state $X(2050) \rightarrow
\Sigma^*(1385)^oK^+$ which was observed in coherent reaction (14) in the
region of very small transverse momenta $(P^2_T < 0.02$~GeV$^2$) is also produced
not by diffractive, but by  the electromagnetic Coulomb production mechanism~[46].

The feasibility to separate the Coulomb production processes in the coherent
proton reactions at $E_p$~= 70~GeV on the carbon target in the measurements with
the SPHINX setup was demonstrated recently by observation of the Coulomb
production of $\Delta(1232)^+$ isobar with $I~= 3/2$ in the reaction
\begin{equation}
p + C \rightarrow \Delta(1232)^+ + C
\end{equation}
(see~[46]).

\section{RELIABILITY OF X(2000) BARYON STATE}

(large decay branching with strange particle emission, limited
decay width) were obtained with a good statistical significance
in  different SPHINX runs with widely different experimental
conditions and for several kinematical regions of reaction (16).
The average values of the mass and width of X(2000) state are
\begin{equation}
X(2000) \rightarrow \Sigma^oK^+
\left \{  \begin{array}{ccc}
M & = & 1989 \pm 6~{\rm{MeV}}\\
\Gamma& = & ~~91 \pm 20~{\rm{MeV}}
\end{array} \right.
\end{equation}

The data on X(2000) baryon state with unusual dynamical properties

Due to its anomalous properties X(2000) state can be considered as a serious
candidate for pentaquark exotic baryon with hidden
strangeness: $|X(2000)> = |uuds \bar s>$. Recently
some new additional data have been
obtained which are in favor of the reality of X(2000) state.

\newpage
\begin{figure}[H] 
\parbox[b]{7.5cm}{
a) In the experiment of the SPHINX Collaboration
reaction (17) was studied. The data for the effective mass spectrum
$M(\Sigma^+K^o)$~ in this reaction are presented in Fig.~12.
In spite of limited statistics the X(2000) peak and the indication for
X(1810) structure are seen in this mass spectrum and are quite compatible
with the data for reaction  (16).

b) In the experiment on the SELEX(E781) spectrometer with the $\Sigma^-$
hyperon beam of the Fermilab Tevatron  the diffractive production reaction
\begin{equation}
\Sigma^- + N \rightarrow [\Sigma^-K^+K^-] + N
\end{equation}
was studied at the beam momentum $P_{\Sigma^-} \simeq 600$~GeV. In the
invariant mass spectrum $M(\Sigma^-K^+)$ for this reaction a peak with
parameters $M = 1962 \pm 12$~MeV and $\Gamma = 96 \pm 32$~MeV 
was observed (Fig.~13).
} \hfill 
\parbox[b]{7.5cm}{
\psfig{file=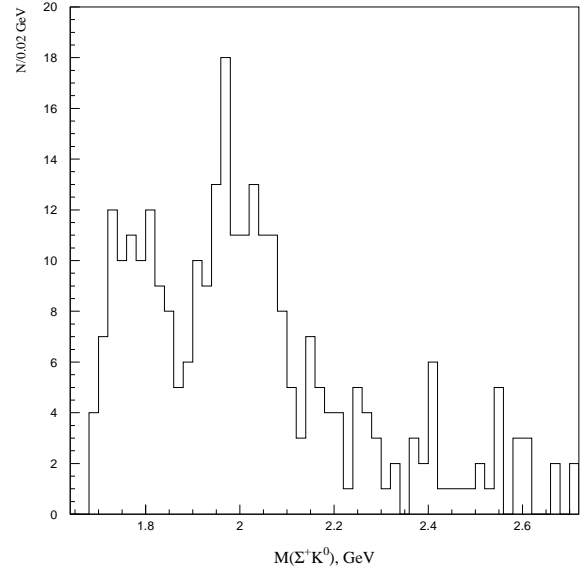,width=7.5cm,clip=}
{\vspace*{-0.2cm}\caption{\small The  effective mass spectrum 
$M(\Sigma^+K^o)$ in reaction (17) for $P^2_T<0.1$~GeV$^2$.}}
}
\end{figure}
\noindent
The parameters of this structure are very near to the parameters of
$X(2000) \rightarrow \Sigma^0K^+$ state which was observed in the experiments
on the SPHINX spectrometer. It seems, that the real existence of $X(2000)$
baryon is supported by the data of another experiment and in another process.

\begin{figure}[H]
\centerline{\psfig{file=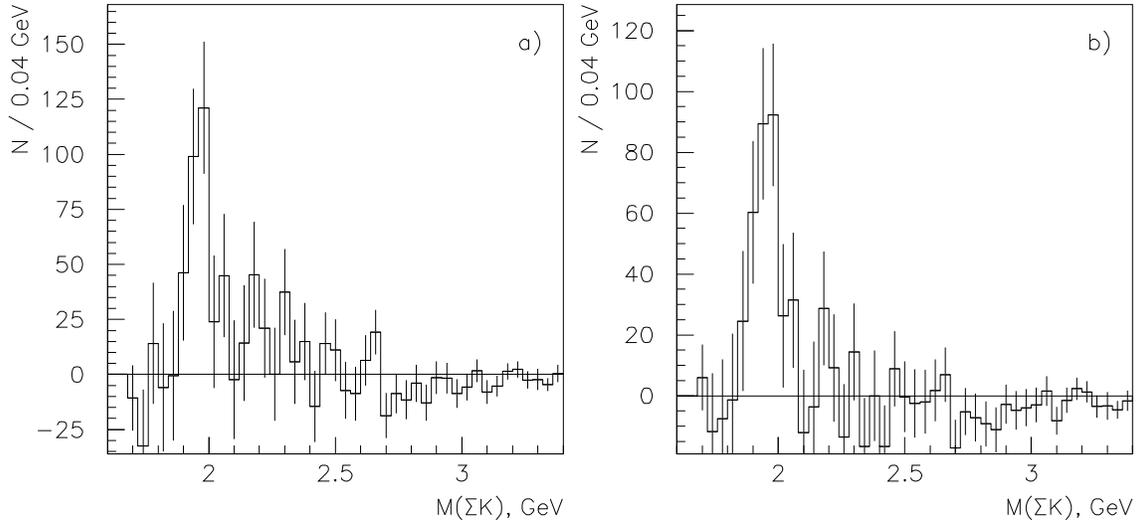,width=15cm,clip=}}
\caption{Study of $M(\Sigma^-K^+)$ in the reaction $\Sigma^- + N \rightarrow
[\Sigma^-K^+K^-] + N$ in the SELEX experiment. Here the spectrum $M(\Sigma^-K^-)$
with open exotic quantum number used for subtraction of nonresonance
background in $M(\Sigma^-K^+)$ after some normalization.One presents in this figure
$M(\Sigma^-K^+) - 0.95 M(\Sigma^-K^-)$ (here 0.95~--- normalization factor):
a) all the events; b) after  subtraction of the events in $\phi$ band
to suppress the influence of the reaction $\Sigma^- + N \rightarrow [\Sigma^-\phi] + N$.}
\end{figure}

Preliminary results of studying reactions (17) and (37) were discussed in the
talks at the last conferences~[24,25,47] and are now under detailed study.

\section{NONPERIPHERAL PROCESSES}

As was discussed above (Section~2.3) the search for new baryons in proton-induced
diffractive-like reactions in the nonperipheral domain, with $P^2_T > 0.3-0.5$~GeV$^2$,
seems to be quite promising. Here we present the very first results of these
searches in the invariant mass spectra of the $\Sigma^oK^+$ and $p\eta$ systems
produced in the reactions $p + N \rightarrow [\Sigma^oK^+] + N$ (16)
and $p + N \rightarrow [p\eta] + N$ (18) for $P^2_T > 0.3$~GeV$^2$
(see~[20,21]).Combined data on reaction (16) from the old and new runs are
shown in Fig.~14a. The data from the old run for reaction (18) are shown in Fig.~14b.
Despite limited statistics, a structure with mass $M~\approx 2350$~MeV and
width $\Gamma~\sim 60$~MeV can be clearly seen in these two mass spectra.
They require a further study in future experiments with large statistics. The
same statement seems true for intriguing data on the invariant mass
spectrum $M(p\eta')$ for reaction (19) in the region $P^2_T > 0.3$~GeV$^2$ (see
Fig.~14c). It must be stressed that reaction (19) is the only one (among more
than a dozen of other diffractive-like reactions studied in the SPHINX experiments)
in which a strong coherent production on carbon nuclei was not observed (the
absence of the forward peak in $dN/dP^2_T$ distribution with the slope value
$b \gtrsim 50$~GeV$^{-2}$; the slope for forward cone in (19) is
$b \sim 6.5$~GeV$^{-2}$~--- see~[20,22]).

\begin{figure}[htb]
\centerline{\psfig{file=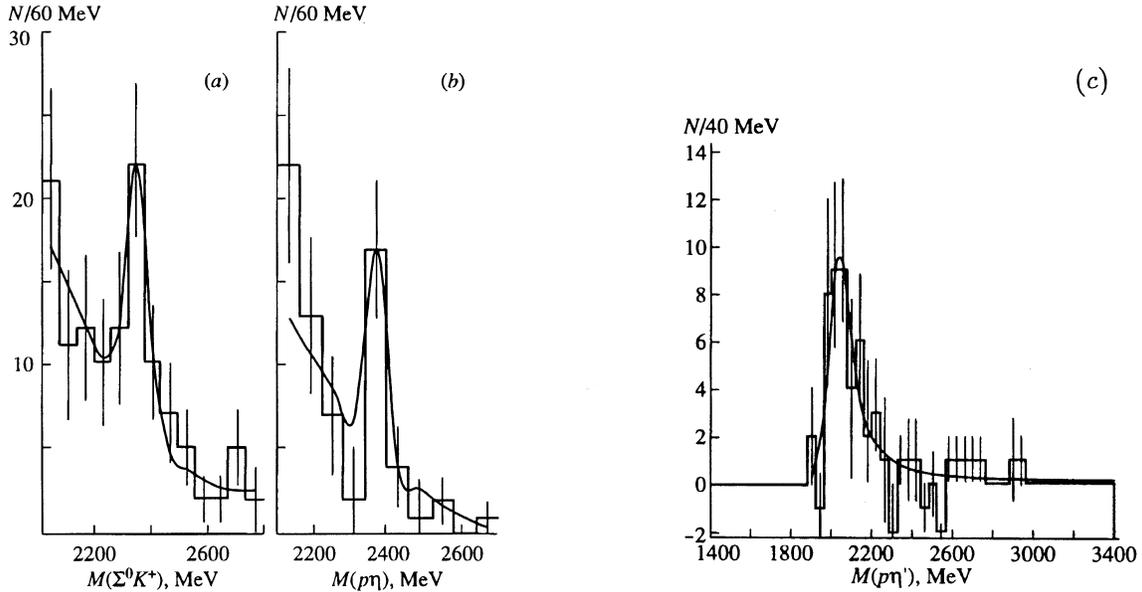,width=15cm,clip=}}
\caption{(a) Invariant mass spectrum of the $\Sigma^oK^+$ system produced in the
reaction $p + N \rightarrow [\Sigma^oK^+] + N$ (16) for $P^2_T > 0.3$~GeV$^2$
(combined data from the old and new runs). (b) Invariant mass spectrum of the
$p\eta$ system produced in the reaction $p + N \rightarrow [p\eta] + N$ (18)
for $P^2_T > 0.3$~GeV$^2$ (old data, see~[20]). A narrow structure with
$M \sim 2350$~MeV and $\Gamma \sim 60$~MeV in the nonperipheral region
of reactions  (16) and (18) is seen in these mass spectra.
(c) Invariant mass spectrum $M(p\eta ')$ for the reaction $p + N \rightarrow
[p\eta '] + N$ (19) in the region $P^2_T > 0.3$~GeV$^2$ (old run).
This spectrum is dominated by a threshold structure with
$M \sim 2000$~MeV and $\Gamma \sim 100$~MeV.}
\end{figure}

\section*{CONCLUSION}

The extensive research program of studying diffractive production in
$E_p = 70$~GeV proton reactions is being carried out in experiments with the
SPHINX setup~[2, 12-28, 45-47]. This program is aimed primarily at searches for
cryptoexotic baryons with hidden strangeness $(|uuds\bar s>)$. Only a part
of these experiments is discussed here.

The most important results of these searches were obtained in studies of the
hyperon-kaon systems produced in proton diffractive dissociation processes and
first of all in reaction $p + N \rightarrow \Sigma^oK^+ + N$~(16).

New data for this diffractive production reaction were obtained with the
partially upgraded SPHINX detector (with new $\gamma$-spectrometer and with
better possibilities to detect $\Lambda \rightarrow p \pi^-$ and $\Sigma^o
\rightarrow \Lambda \gamma$ decays). New data are in a good agreement with
previous SPHINX results on the invariant mass spectrum  $M(\Sigma^oK^+)$
in this reaction.

A strong $X(2000)$ peak with $M = 1989 \pm 6$~GeV and $\Gamma = 91 \pm 20$~MeV
together with a narrow threshold structure (with $M \sim 1810$~MeV and
$\Gamma \sim 60$~MeV) are clearly seen in the $(\Sigma^oK^+)$ invariant mass
spectra. The latter structure is produced at very small transverse momenta,
$P^2_T < 0.01 - 0.02$~GeV$^2$. Unusual properties of the $X(2000)$ baryon
state (narrow decay width, anomalously large branching ratio for the decays
with strange particle emission) make this state  a serious candidate for
a cryptoexotic pentaquark baryon with hidden strangeness $|qqqs\bar s>$.
Preliminary data for $|\Sigma K>$ states in other reactions (17) and (37)
confirm the real existence of $X(2000)$ baryon.

Several other interested phenomena were observed in the nonperipheral
domain (with $P^2_T \gtrsim 0.3$~GeV$^2$), in the study of different reactions
(for example, reactions (14), (18) and (19)). But they need experimental
verification with much better statistics.

Now the first stage of the experimental program on the SPHINX setup has been
completed. In the last years the SPHINX spectrometer was totally upgraded. The
luminosity and the data taking rate were greatly increased. In the recent runs
with this upgraded setup we obtained a large statistics which is now
under  data analysis. In the near future, we expect to increase statistics
by an order of magnitude
for the processes discussed above and for some other proton reactions.

This work is partially supported by Russian Foundation for Basic Research.


\begin{thebibliography}{58}
\bibitem{1}
L.G.Landsberg. Surv. High Energy Phys. 6(1992), 257.
\bibitem{2}
L.G.Landsberg. Yad.Fiz. 57 (1994) 47 [Phys. At. Nucl. (Engl.Transl.)
57 (1994) 42]; USP. Fiz. Nauk. 164 (1994) 1129. [Physics-Uspekhi  (Engl. Transl.)
37 (1994) 1043].
\bibitem{3}
K.Peters. Nucl. Phys. A558 (1993) 92.
\bibitem{4}
C.B.Dover. Nucl. Phys. A558 (1993) 721.
\bibitem{5}
C.Amsler. Rapporter talk. Proc. Conf on High Enerhy Phys. (ICHER), Glasgow, Scotland,
July 1994.
\bibitem{6}
F.E.Close. Preprint RAL-87-072, Chilton, 1987.
\bibitem{7}
P.Blum. Int. J.Mod. Phys. 11 (1996), 3003.
\bibitem{8}
T.Hirose et. al. Nuov. Cim. 50 (1979) 120;
C.Fucunage et al. Nuov. Cim. 58 (1980) 199.
\bibitem{9}
J.Amizzadeh et al. Phys. Lett. B89 (1979) 120.
\bibitem{10}
A.N.Aleev. et al. Z.Phys. C25 (1984) 205.
\bibitem{11}
V.M.Karnaukhov V.M. et al. Phys. Lett. B281 (1992) 148.
\bibitem{12}
D.V.Vavilov et al. (SPHINX Collab.). Yad. Fiz. 57 (1994) 241 [Phys. At. Nucl.
(Engl. Transl.) 57 (1994) 227]; M.Ya.Balatz  et al. (SPHINX Collab.). Z.Phys.
C61 (1994) 220.
\bibitem{13}
M.Ya.Balatz  et al. (SPHINX Collab.). Z.Phys. C61 (1994) 399.
\bibitem{14}
D.V.Vavilov  et al. (SPHINX Collab). Yad. Fiz. 57 (1994) 253 [Phys. At. Nucl.
(Engl. Transl.) 57 (1994) 238].
\bibitem{15}
L.G.Landsberg  et al (SPHINX Collab.). Nuov. Cim. A107 (1994) 2441.
\bibitem{16}
V.F.Kurshetsov, L.G.Landsberg. Yad. Fiz. 57 (1994) 2030 [Phys. At. Nucl.
(Engl. Tranl.) 57 (1994) 1954].
\bibitem{17}
D.V.Vavilov  et al. (SPHINX Collab.). Yad. Fiz. 57 (1994) 1449 [Phys. At. Nucl.
(Engl. Transl.) 57 (1994) 1376].
\bibitem{18}
D.V.Vavilov et al. (SPHINX Collab.). Yad. Fizz. 58 (1995) 1426 [Phys. At.
Nucl. (Engl. Transl.)] 58 (1995) 1342.
\bibitem{19}
S.V.Golovkin  et al. (SPHINX Collab.). Z. Phys. C68 (1995) 585.
\bibitem{20}
S.V.Golovkin  et al. (SPHINX Collab.). Yad. Fiz. 59 (1996) 1395 [Phys. At. Nucl.
(Engl. Transl.). 59 (1996) 1336].
\bibitem{21}
V.A.Bezzubov et al. (SPHINX Collab.). Yad. Fiz. 59 (1996) 2199 [Phys. At.
Nucl. (Engl. Transl.). 59 (1996) 2117].
\bibitem{22}
L.G.Landsberg.  Yad. Fiz. 60 (1997) 1541. [Phys. At. Nucl. (Engl. Transl.). 60
(1997) 1397].
\bibitem{23}
L.G.Landsberg. Hadron Spectroscopy (``Hadron 97''). Seventh Intern. Conf. Upton, NY,
August 1997 (ed. S.-U.Chung, H.J.Willutzki), p. 725.
\bibitem{24}
L.G.Landsberg.  Proc. of 4th Workshop on Small-X and Diffractive Physics,
Fermilab, Batavia, 17-20 September 1998, p.189.
\bibitem{25}
L.G.Landsberg. Plenary Talk on the Conf. ``Fundamental Interactions of
Elementary Particles'', ITEP, Moscow, November 1998. Yad. Fiz. (in press).
\bibitem{26}
V.A.Victorov  et al. Yad. Fiz. 59 (1996) 1229 [Phys. At. Nucl. (Engl. Transl.)
59 (1996) 1175].
\bibitem{27}
M.Ya.Balatz et al. (SPHINX Collab.). Yad. Fiz. 59 (1996) 1242 [Phys. At.
Nucl. (Engl. Transl.) 59 (1996) 1186].
\bibitem{28}
S.V.Golovkin  et al. Z. Phys. A359  (1997) 435.
\bibitem{29}
Hong-Mo Chan and H.Hogaasen. Phys. Lett. B72 (1977) 121; Hong-Mo Chan et al.
Phys. Lett. B76 (1978) 634.
\bibitem{30}
H.Hogaasen  and P.Sorba.  Nucl. Phys. B145 (1978) 119; Invited Talk at
Conf. on Hadron Interactions at High Energy. Marseilles (France) 1978.
\bibitem{31}
M.De Crombrughe  et al. Nucl. Phys. B156 (1979) 347.
\bibitem{32}
Hong-Mo Chan, S.T.Tsou. Nucl. Phys. B118 (1977) 413.
\bibitem{33}
C.Caso  et al. (PDG). The Europ. Phys. Journ., 3 (1998) 1.
\bibitem{34}
L.G.Landsberg.  Usp. Fiz. Nauk. 160 (1990) 1.
\bibitem{35}
G.Bellini  et al. Nuov. Cim. A79 (1984) 282.
\bibitem{36}
D.Alde et al. (GAMS Collab.) Phys. Lett. B182 (1986) 105; Phys. Lett. B276
(1992) 457.
\bibitem{37}
D.Alde  et al. (GAMS Collab). Phys. Lett. B216 (1989) 447; Phys. Lett. B276
(1992) 375.
\bibitem{38}
S.S.Gershtein. Proc. 3rd Int. Conf. on Hadron Spectroscopy: ``Hadron-89''
(Eds. F. Binon et al.) Paris (1989) 175.
\bibitem{39}
R.A.Shumacher. Preprint CMU MEG-96-007, Pittsburg, 1996.
\bibitem{40}
H.Primakoff. Phys. Rev. 81 (1951) 899.
\bibitem{41}
I.Ya.Pomeranchuk  and I.M.Shmushkevitch. Nucl. Phys. B23 (1961) 452.
\bibitem{42}
A.Halpern et al. Phys. Rev. 152 (1966) 1295; J.Dreitlein, H.Primakoff. Phys.
Rev. 125 (1962) 591.
\bibitem{43}
M.Zielinski  et al. Z.Phys. C31 (1986) 545; C34 (1986) 255.
\bibitem{44}
L.G.Landsberg. Nucl. Phys. (Proc. Suppl.) B211 (1991) 179c; Yad. Fiz. 52 (1990) 192.
\bibitem{45}
S.I.Golovkin  et al. Europ. Phys. Journ. A (to be published).
\bibitem{46}
D.V.Vavilov  et al. Yad. Fiz. 62 (1999) (to be published).
\bibitem{47}
G.S.Lomkazi. Talk on the Symposium on Modern Trends in Particles Physics,
dedicated to the 70 anniversary of G.Chikovani. Tbilisi, Georgia, September 1998.
\end{thebibliography}
\end{document}